\input harvmac.tex

\def\ap{\alpha'}

\lref\fzz{V.~Fateev, A.~B.~Zamolodchikov and A.~B.~Zamolodchikov,
``Boundary Liouville field theory. I: Boundary state and boundary
two-point function,'' arXiv:hep-th/0001012.
}

\lref\nati{ G.~W.~Moore and N.~Seiberg, ``From loops to fields in
2-D quantum gravity,'' Int.\ J.\ Mod.\ Phys.\ A {\bf 7}, 2601
(1992).
}

\lref\ponsot{B.~Ponsot and J.~Teschner, ``Boundary Liouville field
theory: Boundary three point function,'' Nucl.\ Phys.\ B {\bf
622}, 309 (2002) [arXiv:hep-th/0110244].
}

\lref\zz{A.~B.~Zamolodchikov and A.~B.~Zamolodchikov, ``Liouville
field theory on a pseudosphere,'' arXiv:hep-th/0101152.
}

\lref\gk{ D.~J.~Gross and I.~R.~Klebanov, ``Fermionic String Field
Theory Of C = 1 Two-Dimensional Quantum Gravity,'' Nucl.\ Phys.\ B
{\bf 352}, 671 (1991).
}

\lref\igor{ I.~R.~Klebanov, ``String theory in two-dimensions,''
arXiv:hep-th/9108019.
}

\lref\dj{ S.~R.~Das and A.~Jevicki, ``String Field Theory And
Physical Interpretation Of D = 1 Strings,'' Mod.\ Phys.\ Lett.\ A
{\bf 5}, 1639 (1990).
}

\lref\jp{ J.~Polchinski, ``Classical Limit Of (1+1)-Dimensional
String Theory,'' Nucl.\ Phys.\ B {\bf 362}, 125 (1991).
}

\lref\mgv{ J.~McGreevy and H.~Verlinde, ``Strings from tachyons:
The c = 1 matrix reloaded,'' arXiv:hep-th/0304224.
}

\lref\llm{ N.~Lambert, H.~Liu and J.~Maldacena, ``Closed strings
from decaying D-branes,'' arXiv:hep-th/0303139.
}

\lref\BPIZ{ E. Brezin, C. Itzykson, G. Parisi and J. Zuber,
CMP {\bf 59} (1978) 35.}

\lref\joe{J.~Polchinski, ``What is String Theory?''
arXiv:hep-th/9411028
}

\lref\ginsparg{ P.~Ginsparg and G.~W.~Moore, ``Lectures On 2-D
Gravity And 2-D String Theory,'' arXiv:hep-th/9304011.
}

 \lref\Pol{ A.~M.~Polyakov, ``Gauge fields and space-time,''
Int.\ J.\ Mod.\ Phys.\ A {\bf 17S1}, 119 (2002)
[arXiv:hep-th/0110196].
}

\lref\shenker{ S.~H.~Shenker, ``The Strength Of Nonperturbative
Effects In String Theory,'' RU-90-47
{\it Presented at the Cargese Workshop on Random Surfaces,
Quantum Gravity and Strings, Cargese, France, May 28 - Jun 1, 1990}
}

\lref\gaiotto{ D.~Gaiotto, N.~Itzhaki and L.~Rastelli, ``Closed
strings as imaginary D-branes,'' arXiv:hep-th/0304192.
}

\lref\Polch{ J.~Polchinski, ``Combinatorics Of Boundaries In
String Theory,'' Phys.\ Rev.\ D {\bf 50}, 6041 (1994)
[arXiv:hep-th/9407031].
}

 \lref\Green{ M.~B.~Green, ``A Gas of D instantons,'' Phys.\
Lett.\ B {\bf 354}, 271 (1995) [arXiv:hep-th/9504108].
}

\lref\SenNX{ A.~Sen and B.~Zwiebach, ``Tachyon condensation in
string field theory,'' JHEP {\bf 0003}, 002 (2000)
[arXiv:hep-th/9912249].
}

 \lref\gregk{ D.~Kutasov, M.~Marino and G.~W.~Moore, ``Some exact
results on tachyon condensation in string field theory,'' JHEP
{\bf 0010}, 045 (2000) [arXiv:hep-th/0009148].
}

 \lref\GerasimovZP{ A.~A.~Gerasimov and S.~L.~Shatashvili, ``On
exact tachyon potential in open string field theory,'' JHEP {\bf
0010}, 034 (2000) [arXiv:hep-th/0009103].
}

\lref\GutperleAI{
M.~Gutperle and A.~Strominger,
``Spacelike branes,''
JHEP {\bf 0204}, 018 (2002)
[arXiv:hep-th/0202210].
}

\lref\PolchinskiJP{ J.~Polchinski, ``On the nonperturbative
consistency of d = 2 string theory,'' Phys.\ Rev.\ Lett.\  {\bf
74}, 638 (1995) [arXiv:hep-th/9409168].
}

\lref\ZwiebachIE{
B.~Zwiebach,
``Closed string field theory: Quantum action and the B-V master equation,''
Nucl.\ Phys.\ B {\bf 390}, 33 (1993)
[arXiv:hep-th/9206084].
}

\lref\DiVecchiaPR{
P.~Di Vecchia, M.~Frau, I.~Pesando, S.~Sciuto, A.~Lerda and R.~Russo,
``Classical p-branes from boundary state,''
Nucl.\ Phys.\ B {\bf 507}, 259 (1997)
[arXiv:hep-th/9707068].
}

\lref\zuber{
C.~Itzykson and J.~B.~Zuber,
``Quantum Field Theory,''
}

\lref\GarousiTR{
M.~R.~Garousi,
``Tachyon couplings on non-BPS D-branes and Dirac-Born-Infeld action,''
Nucl.\ Phys.\ B {\bf 584}, 284 (2000)
[arXiv:hep-th/0003122].
}

\lref\SenMD{
A.~Sen,
``Supersymmetric world-volume action for non-BPS D-branes,''
JHEP {\bf 9910}, 008 (1999)
[arXiv:hep-th/9909062].
}

\lref\MoellerVX{
N.~Moeller and B.~Zwiebach,
``Dynamics with infinitely many time derivatives and rolling tachyons,''
JHEP {\bf 0210}, 034 (2002)
[arXiv:hep-th/0207107].
}

\lref\gsw{
M.~B.~Green, J.~H.~Schwarz and E.~Witten,
``Superstring Theory. Vol. 1: Introduction,''}

\lref\SenNU{ A.~Sen, ``Rolling tachyon,'' JHEP {\bf 0204}, 048
(2002) [arXiv:hep-th/0203211].
}

\lref\SenIN{ A.~Sen, ``Tachyon matter,'' JHEP {\bf 0207}, 065
(2002) [arXiv:hep-th/0203265].
}

\lref\SenTM{ A.~Sen, ``Dirac-Born-Infeld Action on the Tachyon
Kink and Vortex,'' [arXiv:hep-th/0303057].
}

\lref\SenQA{ A.~Sen, ``Time and tachyon,'' [arXiv:hep-th/0209122].
}

\lref\SenAN{ A.~Sen, ``Field theory of tachyon matter,'' Mod.\
Phys.\ Lett.\ A {\bf 17}, 1797 (2002) [arXiv:hep-th/0204143].
}

\lref\SenVV{ A.~Sen, ``Time evolution in open string theory,''
JHEP {\bf 0210}, 003 (2002) [arXiv:hep-th/0207105].
}

\lref\ChenFP{ B.~Chen, M.~Li and F.~L.~Lin, ``Gravitational
radiation of rolling tachyon,'' JHEP {\bf 0211}, 050 (2002)
[arXiv:hep-th/0209222].
}

\lref\SenMS{
P.~Mukhopadhyay and A.~Sen,
``Decay of unstable D-branes with electric field,''
JHEP {\bf 0211}, 047 (2002)
[arXiv:hep-th/0208142].
}

\lref\Rey{S.~J.~Rey and S.~Sugimoto,
``Rolling Tachyon with Electric and Magnetic Fields -- T-duality approach,''
[arXiv:hep-th/0301049].
}

\lref\GutperleAI{ M.~Gutperle and A.~Strominger, ``Spacelike
branes,'' JHEP {\bf 0204}, 018 (2002) [arXiv:hep-th/0202210].
}

\lref\StromingerPC{ A.~Strominger, ``Open string creation by
S-branes,'' [arXiv:hep-th/0209090].
}

\lref\GutperleBL{M.~Gutperle and A.~Strominger,
``Timelike Boundary Liouville Theory,''
[arXiv:hep-th/0301038].
}

\lref\WaldWT{
R.~M.~Wald,
``Existence Of The S Matrix In Quantum Field Theory In Curved Space-Time,''
Annals Phys.\  {\bf 118}, 490 (1979).
}

\lref\Larsen{F.~Larsen, A.~Naqvi and S.~Terashima,
``Rolling tachyons and decaying branes,''
[arXiv:hep-th/0212248].
}
\lref\krauslarsen{
B.~Craps, P.~Kraus and F.~Larsen,
``Loop corrected tachyon condensation,''
JHEP {\bf 0106}, 062 (2001)
[arXiv:hep-th/0105227].
}

\lref\weinberg{
S. Weinberg,  ``Gravitation and Cosmology''.
}

\lref\Callan{
C.~G.~Callan, I.~R.~Klebanov, A.~W.~Ludwig and J.~M.~Maldacena,
``Exact solution of a boundary conformal field theory,''
Nucl.\ Phys.\ B {\bf 422}, 417 (1994)
[arXiv:hep-th/9402113].
}

 \lref\larus{ J.~Polchinski and L.~Thorlacius, ``Free Fermion
Representation Of A Boundary Conformal Field Theory,'' Phys.\
Rev.\ D {\bf 50}, 622 (1994) [arXiv:hep-th/9404008].
}

 \lref\teschner{ J.~Teschner, ``Remarks on Liouville theory with
boundary,'' arXiv:hep-th/0009138.
}

\lref\RS{
A.~Recknagel and V.~Schomerus,
``Boundary deformation theory and moduli spaces of D-branes,''
Nucl.\ Phys.\ B {\bf 545}, 233 (1999)
[arXiv:hep-th/9811237].
}

\lref\SenMG{
A.~Sen,
``Non-BPS states and branes in string theory,''
[arXiv:hep-th/9904207].
}

\lref\juan{J.~Maldacena, Unpublished}

\lref\shiraz{S.~Minwalla and K.~Pappododimas, Unpublished}

\lref\hwang{
S.~Hwang,
``Cosets as gauge slices in SU(1,1) strings,''
Phys.\ Lett.\ B {\bf 276}, 451 (1992)
[arXiv:hep-th/9110039];
J.~M.~Evans, M.~R.~Gaberdiel and M.~J.~Perry,
``The no-ghost theorem for AdS(3) and the stringy exclusion principle,''
Nucl.\ Phys.\ B {\bf 535}, 152 (1998)
[arXiv:hep-th/9806024].
}

\lref\polchinski{
J.~Polchinski,
``String Theory. Vol. 1: An Introduction To The Bosonic String,''
}

\lref\GurarieXQ{
V.~Gurarie,
``Logarithmic operators in conformal field theory,''
Nucl.\ Phys.\ B {\bf 410}, 535 (1993)
[arXiv:hep-th/9303160].
}

\lref\fs{
W.~Fischler and L.~Susskind,
``Dilaton Tadpoles, String Condensates And Scale Invariance,''
Phys.\ Lett.\ B {\bf 171}, 383 (1986);
``Dilaton Tadpoles, String Condensates And Scale Invariance. 2,''
Phys.\ Lett.\ B {\bf 173}, 262 (1986).
}

\lref\MaloneyCK{
A.~Maloney, A.~Strominger and X.~Yin,
``S-brane thermodynamics,''
[arXiv:hep-th/0302146].
}

\lref\BuchelTJ{
A.~Buchel, P.~Langfelder and J.~Walcher,
``Does the tachyon matter?,''
Annals Phys.\  {\bf 302}, 78 (2002)
[arXiv:hep-th/0207235];
A.~Buchel and J.~Walcher,
``The tachyon does matter,''
arXiv:hep-th/0212150.
}

\lref\LeblondDB{ F.~Leblond and A.~W.~Peet,
``SD-brane gravity fields and rolling tachyons,''
[arXiv:hep-th/0303035].
}

\lref\FischlerJA{
W.~Fischler, S.~Paban and M.~Rozali,
``Collective Coordinates for D-branes,''
Phys.\ Lett.\ B {\bf 381}, 62 (1996)
[arXiv:hep-th/9604014];
``Collective coordinates in string theory,''
Phys.\ Lett.\ B {\bf 352}, 298 (1995)
[arXiv:hep-th/9503072].
}

\lref\PO{
V.~Periwal and O.~Tafjord,
``D-brane recoil,''
Phys.\ Rev.\ D {\bf 54}, 3690 (1996), arXiv:hep-th/9603156.
}

\lref\KutasovER{ D.~Kutasov and V.~Niarchos, ``Tachyon effective
actions in open string theory,'' arXiv:hep-th/0304045.
}

\lref\OkuyamaWM{ K.~Okuyama, ``Wess-Zumino term in tachyon
effective action,'' arXiv:hep-th/0304108.
}

\lref\KMW{
J.~S.~Caux, I.~I.~Kogan and A.~M.~Tsvelik,
``Logarithmic Operators and Hidden Continuous Symmetry in Critical Disordered Models,''
Nucl.\ Phys.\ B {\bf 466}, 444 (1996)
[arXiv:hep-th/9511134];
I.~I.~Kogan and N.~E.~Mavromatos,
``World-Sheet Logarithmic Operators and Target Space Symmetries in String Theory,''
Phys.\ Lett.\ B {\bf 375}, 111 (1996)
[arXiv:hep-th/9512210];
I.~I.~Kogan, N.~E.~Mavromatos and J.~F.~Wheater,
``D-brane recoil and logarithmic operators,''
Phys.\ Lett.\ B {\bf 387}, 483 (1996),
arXiv:hep-th/9606102.
}

\lref\FlohrZS{
M.~Flohr,
``Bits and pieces in logarithmic conformal field theory,''
arXiv:hep-th/0111228.
}

\lref\GaberdielTR{
M.~R.~Gaberdiel,
``An algebraic approach to logarithmic conformal field theory,''
arXiv:hep-th/0111260.
}

\lref\OkudaYD{ T.~Okuda and S.~Sugimoto, ``Coupling of rolling
tachyon to closed strings,'' Nucl.\ Phys.\ B {\bf 647}, 101 (2002)
[arXiv:hep-th/0208196].
}

\lref\Aref{
I.~Y.~Aref'eva, L.~V.~Joukovskaya and A.~S.~Koshelev,
``Time evolution in superstring field theory on non-BPS brane.
I: Rolling  tachyon and energy-momentum
conservation,''
arXiv:hep-th/0301137.
}

\lref\IshidaCJ{
A.~Ishida and S.~Uehara,
``Rolling down to D-brane and tachyon matter,''
JHEP {\bf 0302}, 050 (2003)
[arXiv:hep-th/0301179].
}

\lref\KlusonAV{
J.~Kluson,
``Time dependent solution in open Bosonic string field theory,''
arXiv:hep-th/0208028;
``Exact solutions in open Bosonic string field theory and
marginal  deformation in CFT,''
[arXiv:hep-th/0209255].
}

\lref\MinahanIF{
J.~A.~Minahan,
``Rolling the tachyon in super BSFT,''
JHEP {\bf 0207}, 030 (2002)
[arXiv:hep-th/0205098].
}

\lref\SugimotoFP{
S.~Sugimoto and S.~Terashima,
``Tachyon matter in boundary string field theory,''
JHEP {\bf 0207}, 025 (2002)
[arXiv:hep-th/0205085].
}

\lref\ReyXS{
S.~J.~Rey and S.~Sugimoto,
``Rolling tachyon with electric and magnetic fields: T-duality approach,''
arXiv:hep-th/0301049.
}

\lref\rastelli{  D.~Gaiotto, N.~Itzhaki and L.~Rastelli, ``Closed
strings as imaginary D-branes,'' arXiv:hep-th/0304192.
}

\lref\LambertHK{
N.~D.~Lambert and I.~Sachs,
``Tachyon dynamics and the effective action approximation,''
Phys.\ Rev.\ D {\bf 67}, 026005 (2003)
[arXiv:hep-th/0208217].
}

\lref\HarveyQU{ J.~A.~Harvey, P.~Horava and P.~Kraus,
``D-sphalerons and the topology of string configuration space,''
JHEP {\bf 0003}, 021 (2000) [arXiv:hep-th/0001143].
}

\lref\MooreIR{ G.~W.~Moore, N.~Seiberg and M.~Staudacher, ``From
loops to states in 2-D quantum gravity,'' Nucl.\ Phys.\ B {\bf
362}, 665 (1991).
}

\lref\DrukkerWX{ N.~Drukker, D.~J.~Gross and N.~Itzhaki,
``Sphalerons, merons and unstable branes in AdS,'' Phys.\ Rev.\ D
{\bf 62}, 086007 (2000) [arXiv:hep-th/0004131].
}

\lref\yang{ Z.~Yang, ``Dynamical Loops In D = 1 Random Matrix
Models,'' Phys.\ Lett.\ B {\bf 257}, 40 (1991).
}

\lref\min{ J.~A.~Minahan, ``Matrix models and one-dimensional open
string theory,'' Int.\ J.\ Mod.\ Phys.\ A {\bf 8}, 3599 (1993)
[arXiv:hep-th/9204013].
}

\lref\kk{ V.~A.~Kazakov and I.~K.~Kostov, ``Loop gas model for
open strings,'' Nucl.\ Phys.\ B {\bf 386}, 520 (1992)
[arXiv:hep-th/9205059].
}
\lref\sengupta{
A.~M.~Sengupta and S.~R.~Wadia,
``Excitations And Interactions In D = 1 String Theory,''
Int.\ J.\ Mod.\ Phys.\ A {\bf 6}, 1961 (1991).
}
\lref\wad{
A.~Dhar, G.~Mandal and S.~R.~Wadia,
``A Time dependent classical solution of c = 1 string field theory and nonperturbative effects,''
Int.\ J.\ Mod.\ Phys.\ A {\bf 8}, 3811 (1993)
[arXiv:hep-th/9212027].
}


\newbox\tmpbox\setbox\tmpbox\hbox{\abstractfont
}

 \Title{\vbox{\baselineskip12pt\hbox to\wd\tmpbox{\hss PUPT-2085
} } } {\vbox{\centerline{D-brane Decay in Two-Dimensional
 String Theory}
}}

\smallskip
\centerline{Igor R. Klebanov,\foot{On leave from
Physics Department, Princeton University.}
Juan Maldacena and Nathan  Seiberg}
\medskip

\smallskip

\centerline{\it Institute for Advanced Study}
 \centerline{\it
Princeton, New Jersey, 08540, USA}

\smallskip

\vglue .3cm

\bigskip
\noindent We consider unstable D0-branes of two dimensional string
theory, described by the boundary state of Zamolodchikov and
Zamolodchikov [hep-th/0101152] multiplied by the Neumann boundary
state for the time coordinate $t$. In the dual description in
terms of the $c=1$ matrix model, this D0-brane is described by a
matrix eigenvalue on top of the upside down harmonic oscillator
potential. As suggested by McGreevy and Verlinde [hep-th/0304224],
an eigenvalue rolling down the potential describes D-brane decay.
As the eigenvalue moves down the potential to the asymptotic
region it can be described as a free relativistic fermion.
Bosonizing this fermion we get a description of the state in terms
of a coherent state of the tachyon field in the asymptotic region,
up to a non-local linear field redefinition by an energy-dependent
phase. This coherent state agrees with the exponential of the
closed string one-point function on a disk with Sen's marginal
boundary interaction for $t$ which describes D0-brane decay.

\Date{May 2003}
%



\newsec{Introduction}

In string theory there are various known cases of unstable
D-branes containing open string tachyon modes on their world
volume; for example, the D-branes of 26-dimensional bosonic
string. In a series of seminal papers Sen proposed a boundary
conformal field theory (BCFT) description of the decay of unstable
D-branes ~\refs{\SenNU\SenIN-\SenAN}. In the bosonic string
theory, the BCFT contains a boundary interaction \eqn\inter{
 \lambda \int d\tau \cosh [X^0 (\tau)/\sqrt{\ap}]\ ,
}
whose exact marginality may be argued via
continuation from the conformal invariance
\refs{\Callan,\larus} of the Euclidean boundary Sine-Gordon theory. A
more detailed study of D$p$-brane decay was carried out in
\refs{\GutperleAI\SenVV\SenMS\StromingerPC\SenQA\Larsen\GutperleBL
\MaloneyCK\ChenFP\Rey\MoellerVX\SugimotoFP\MinahanIF\KlusonAV
\OkudaYD\Aref\IshidaCJ\llm-\gaiotto} and led to many new insights.
A number of
puzzles remain however; for example, to leading order in $g_{s}$
the number of closed string produced in D0-brane decay is UV
divergent and one needs to go beyond tree level for handling
this divergence.

In a thought-provoking recent paper \mgv, a new laboratory for
resolving these puzzles was proposed by J. McGreevy and H.
Verlinde, the 2-dimensional bosonic string theory (for reviews,
see \refs{\igor\ginsparg-\joe}). In $D=2$, the closed string
tachyon becomes massless, so that the theory has no perturbative
instability. The theory is not translation-invariant in the
``Liouville direction'' $\phi$, with the string coupling varying
as $g_s (\phi) \sim e^{2\phi/\sqrt{\ap}}$. The 2-d string theory
was explored intensively in the late 80's and early 90's due in
part to its dual formulation in terms of quantum mechanics of a
large $N$ hermitian matrix often referred to as the $c=1$ matrix
model. The exact solvability of this large $N$ model (it reduces
to $N$ matrix eigenvalues that behave as free fermions \BPIZ)
yielded a wealth of perturbative information about the theory,
some of which has been reproduced using Liouville theory methods.
Furthermore, the model has provided an important hint \shenker\
that stringy non-perturbative effects are of order $e^{-A/g_s}$.
{}From the point of view of the matrix model, such effects are due
to single eigenvalue tunnelling \shenker\ (see also \wad). This observation
preceded the understanding that in string theories such
$O(e^{-A/g_s})$ effects are provided by D-instantons
\refs{\Polch,\Green}. A more general connection between single
eigenvalues in the matrix model and D-branes in the 2-d string
theory was recently proposed in \mgv.  These authors examined a
single eigenvalue balancing at the top of the upside down harmonic
oscillator potential, and studied its decay by rolling down the
potential.

There has also been important recent progress in understanding
D-branes of Liouville theory. In fact, as we review in the next
section, this theory contains different types of D-branes
\refs{\fzz\teschner-\zz}. The D-branes studied in
\refs{\fzz,\teschner} (and \mgv ) are extended in the Liouville
direction and, after inclusion of the time direction, should be
thought of as D1-branes. They are stable. There are also unstable
D-branes localized at large $\phi$ that were discovered and
studied by A.B. and Al.B. Zamolodchikov \zz. After inclusion of
the time direction of the 2-d string, these are D0-branes. They
are remarkable in that the open string theory that lives on them
is {\it exactly} equivalent to quantum mechanics of a tachyon mode
(the only oscillator excitation of the open string gives a
non-dynamical gauge field which implements a constraint on the
wave functions).
The importance of the localized D-branes of \zz\ was emphasized
and the question of their matrix model description was raised in
\Pol.

We suggest that the unstable single eigenvalue of \mgv\ should be
identified with the unstable D0-brane of \zz.  This D-sphaleron
picture for the D0-brane fits nicely with the single-eigenvalue
tunnelling description of the D-instanton anticipated in
\refs{\shenker,\Polch}, and also with the general D-sphaleron
interpretation of unstable D-branes in \HarveyQU (see also
\DrukkerWX ).

More generally, we suggest that the D0-branes dual to the matrix
eigenvalues are the unstable D0-branes of \zz. This provides an
exact duality between open strings on D-branes and closed strings
\mgv.

In the present paper we calculate the decay amplitude of a
D0-brane into closed strings. We show that, to leading order in
$g_s$ the matrix model result agrees with that in Liouville theory
(our results, both in the matrix model and in Liouville theory,
differ from those in \mgv). This leading order rate diverges in
the UV in the same way as in the higher-dimensional calculation of
\llm, but the matrix model approach demonstrates how to regulate
this divergence. In the matrix model the precise regularization
depends on the initial wavefunction we choose for the eigenvalue.
Thus, the matrix model once again seems to provide important
intuition about the quantum behavior of string theory.

\newsec{ D-branes in 2d string theory}

In this section we discuss the different D-branes that exist in
two dimensional string theory.

We are interested in a 2d CFT which is the product of  standard
flat time direction and Liouville theory with $c_L =25$.
Let us first collect formulae and conventions
 about Liouville theory.\foot{ Our conventions are related
to the ones in \refs{\fzz,\zz}  by $
\alpha^{here} = 2 \alpha^{there} $,
$\sqrt{\alpha'} P^{here}  = 2 P^{there}$.}
\eqn\formconv{ c_L = 1 + 6 Q^2 ~,~~~~~~ Q = b + 1/b ~,~~~~~~
\Delta = { 1\over 4} \alpha ( 2 Q - \alpha), ~,~~~~ V = e^{
 \alpha \phi/ \sqrt{\alpha'} } }
For $b=1$, we have $Q=2$, $c_L =25$. The delta function
normalizable states have
 \eqn\norm{ \alpha = Q  + i \sqrt{\alpha'}
 P ~, ~~~~~ \Delta = { Q^2 \over 4 } + {\alpha' \over 4} P^2 }
with real $P$.

When $b \to 1$ the amplitudes in Liouville theory diverge. In order
to obtain finite amplitudes we must also
 send $\mu_0 \to \infty$ keeping
\eqn\keepfin{
\mu \equiv \pi \mu_0 \gamma(b^2)  = ~  {\rm finite} ~, ~~~~~~
}
where $\gamma(b^2) = \Gamma(b^2)/\Gamma(1-b^2) $.
The dependence on $\mu_0$ of the Liouville expressions in \zz , \fzz ,
is only in the combination that remains finite as in \keepfin , so
that we obtain finite expressions. The parameter $\mu$ will be
identified with the matrix model fermi energy.

\subsec{ Extended  D-branes in Liouville theory -- D1-branes}

There are two types of D-branes that have been described in
Liouville theory. First we have D-branes which are extended along
the Liouville direction that were studied in
\refs{\fzz,\teschner,\ponsot}. For these D-branes we can add a
boundary cosmological constant $\mu_B$. The one-point function for
a bulk operator $e^{ \alpha \phi/ \sqrt{\alpha'}}$ is given by
 \eqn\oneext{
U_\alpha = ( \pi \mu_0 \gamma(b^2) )^{ - {i \sqrt{\alpha'} P \over
2b} } \Gamma( 1 + i \sqrt{\alpha'} b P) \Gamma(1 + i
\sqrt{\alpha'} P/b) {2  \cos(  \pi s \sqrt{\alpha'} P) \over i
 \sqrt{\alpha'} P} }
 where $s$ is a parameter parameterizing the
boundary state related to the boundary cosmological constant by
\eqn\bound{ \cosh^2 {\pi s  b}  = {  \mu_{0,B}^2 \over \mu_0 }
\sin {\pi b^2 } } Note that $s$ could be real or imaginary
depending of whether the right hand side of \bound\ is bigger or
smaller than one.

These are the formulae for general $b$.
As we remarked above the formulae for $b=1$ can be obtained by taking
the limit \keepfin . In this case it is also useful to rescale
$ \mu_{0,B}$ in the same way as $\mu_0$ (as in \keepfin ).
 In that way we obtain a finite expression
in the right hand side of \bound\ for $b=1$. So we conclude that
also for $b=1$ we have a continuous family of boundary states
parametrized by $s$. All D-branes in this family have a continuous
spectrum of open string states; they are the open string
``tachyons'' which are massless (as are the closed string
``tachyons''). In particular we can set this rescaled $\mu_B =0$.
This gives us $s = i/2 $. This is just one particular D-brane
state out of this continuous family. This is the D-brane
considered by \mgv .

For $b=1$, the one point function \oneext\ exhibits poles at
imaginary values of $P=in/\sqrt{\alpha'}$ with integer $n$.  The
meaning of these poles is well understood in the matrix model
literature \refs{\igor,\ginsparg}.  For such values of $P$ the
$\phi$ charge of $U_\alpha$ can be screened with an integer number
of insertions of the worldsheet cosmological constant.  Then the
correlation function diverges due to the volume of the noncompact
$\phi$ direction.  These divergences signal the fact that the
correlation function is dominated by the bulk of space time -- the
region $\phi\rightarrow -\infty$.  Therefore these poles signal
that these D-branes are extended in the $\phi $ direction.

\subsec{ Localized D-branes in Liouville -- D0-branes}

There is a second class of Liouville theory
D-branes that were considered in \zz .
These D-branes should be thought of as localized in the Liouville
direction at $\phi =\infty$, which is the strong coupling end.
The localized D-branes are parametrized by two integers $(n,m)$. We are
interested in the case $(n,m) =(1,1)$. The other ones contain
operators more tachyonic than the tachyon in the open string spectrum.
It is not clear what they correspond to.

The bulk one point function in this case reads (eq. (5.10) of \zz)
\eqn\local{
{\cal A}_L \sim
 ( \pi \mu_0 \gamma(b^2) )^{-i {\sqrt{\alpha'}P \over 2 b}}
{ \pi i \sqrt{\alpha'} P
\over
\Gamma( 1 - i \sqrt{\alpha'} b P) \Gamma(1 - i \sqrt{\alpha'} P/b ) }
}
Taking the $b=1$ limit as described above
we get a simple finite answer
\eqn\newlocal{ {\cal A}_L = { 2 \over \sqrt{\pi} } i
\sinh (\pi\sqrt{\alpha'} P)
\mu^{-i {\sqrt{\alpha'} P \over 2}} { \Gamma( i\sqrt{\alpha'} P)
\over \Gamma(- i \sqrt{\alpha'} P )}
\ .
}
where the overall normalization is computed in Appendix B and it
differs from the one in \zz\ due to our conventions.

It is interesting to note that unlike \oneext\ the one point
function \newlocal\ does not exhibit ``bulk poles'' at imaginary
$P$.  Instead, it is zero at $P=-in/\sqrt{\alpha'}$.   We
interpreted the absence of poles as due to the fact that these
D-branes are localized far from the bulk region of large and
negative $\phi$.

\newsec{ D0-brane decay in 2d string theory}

Let us  consider the D0-brane which is the direct product
of the localized Liouville brane times a Neumann boundary condition in
time.
The Liouville part contains only one Virasoro primary, the identity
operator \zz . This is the boundary identity operator and it is
normalizable since the brane is localized in the $\phi$ direction.
It follows that the full physical open string
 spectrum contains only two states, one is
the tachyon and the other is $\partial t$, the gauge field on the
D-brane. This is a negative norm state which is responsible for
imposing the constraint that, when we have $N$ D0-branes of this
type, we should restrict to $U(N)$ invariant wave functions.

Now we can consider the boundary state which is a  product of
Sen's rolling tachyon boundary state for $t$ and the localized
boundary state of \zz\ for $\phi$. We compute the closed string
emission from this state as was done in \llm. We consider a
decaying brane with open string tachyon $T = \lambda \cosh (t/
\sqrt{\alpha'})  $. We get that the decay rate to a massless bulk
tachyons with momentum $P$ and energy $E=|P|$ is \eqn\closedem{
{\cal A} = {\cal A}_t(E) {\cal A}_L(P) } where ${\cal A}_L $ was
defined in \newlocal\ and the time part was computed in \llm\
\eqn\timepart{ {\cal A}_t  =  { \pi e^{- {i } \sqrt{\ap} E \log
\hat \lambda} \over \sinh (\pi \sqrt{\ap} E) } } where $\hat
\lambda = \sin \pi \tilde \lambda $  and  $\tilde \lambda $ is the
variable that appears in the description of the boundary state
\SenNU .

We finally get that the emission rate is
\eqn\abs{
{\cal A} =
 i 2 \sqrt{\pi} {   \sinh (\pi \sqrt{\ap} P)
\over  \sinh (\pi \sqrt{\ap} E) }
e^{ - {i} \sqrt{\ap} E \log \hat \lambda}
\mu^{ - {i\over 2}\sqrt{\ap} P}
{ \Gamma( i \sqrt{\ap} P) \over \Gamma(- i \sqrt{\ap} P )}
 = 2 \sqrt{\pi} i e^{ - {i} \sqrt{\ap} E \log \hat \lambda}
e^{i \delta(P)}
}
where the normalization is computed in the Appendix B.
Remarkably, after we use the on-shell condition
$E=|P|$, we find that the amplitude is a constant
times an
energy dependent phase.\foot{
A part of this phase,
$ e^{i \delta(P)} \equiv \mu^{ - {i\over 2}\sqrt{\ap} P}
{ \Gamma( i \sqrt{\ap} P) \over \Gamma(- i \sqrt{\ap} P )}$
is the standard leg factor that appears for bulk Liouville theory
operators
(see, for instance, \nati,\igor). We discuss this in appendix A.}
As a result,
we find that the rate for the number of particles
$N = \int { dP \over 2 E}|{\cal A}|^2 $ diverges logarithmically, and
the expectation value for the emitted energy diverges linearly in the
UV. This expression also has an IR divergence that is related to the
fact that we are dealing with a massless field in $1+1$ dimensions.

The UV properties are identical to the ones found in \llm\ for the
decay of D0 branes in 26-d bosonic string. In that case, the
exponential decay of the amplitude was exactly cancelled by the
exponentially growing density of closed string states. In the 2-d
string there is no such growth in the density of states. But the
coupling to the tachyon \abs\ does not decay exponentially. In
fact it has a constant absolute value, leading to the same UV
behavior of the particle production rate. This behavior of the
amplitude is consistent with the divergence present in the open
string channel when we consider a D-instanton array (the $\tilde
\lambda =1/2$ states).

Note that the final state that is produced is a coherent state
of the form
\eqn\state{
 |\psi \rangle \sim e^{ \int_0^\infty
 { dp \over \sqrt{2E} } a^\dagger_p
{\cal A} }
|0\rangle
}
up to an overall normalization factor,
where ${\cal A}$ was given
in \abs .

We should end by saying that this does not agree with the
closed string emission formula computed in \mgv , since \mgv\ considered
a D-brane extended in the Liouville direction.

\newsec{The $c=1$ Matrix Model }

Let us recall a few basic facts about the $c=1$ matrix model which
describes two-dimensional string theory. One starts with the
Euclidean path integral for a Hermitian $N\times N$ matrix
\eqn\EQM{ Z=\int D^{N^2}\Phi(x)\exp \biggl [-\beta\int_{-\infty
}^{\infty }dx~\Tr \left (\half (D_x \Phi )^2+ U(\Phi)\right
)\biggr ]\ . }
 where $x$ is the Euclidean time and $\beta$ is the
inverse Planck constant. If we take the matrix potential to be
$U={1\over 2\ap}\Phi^2-{1\over 3}\Phi^3$ then the Feynman graphs
may be thought of as discretized random surfaces embedded into one
Euclidean dimension $x$. It can be shown \igor\ that $\ap$ is the
conventionally normalized inverse string tension parameter (this
follows, for example, from the correct position of poles in the
closed string amplitudes).

{}From the point of view of random surfaces, the matrix $\Phi$ is
an auxiliary concept. Recently, a more direct physical
interpretation of $\Phi$ was proposed \mgv. According to \mgv,
$\Phi$ is an open string tachyon field on D0-branes. We believe
that the D0-branes which this field resides on are obtained from
the localized boundary state for $\phi$ \zz\ multiplied by the
Neumann boundary state for the time coordinate $t=ix$. It follows
that $U(\Phi)$ should be thought of as the tachyon potential.
Indeed, expanding the potential near the quadratic maximum, we
find \eqn\maxt{ U (\Phi) = -{1\over 2\alpha'} (\Phi- \Phi_0)^2 +
O[(\Phi- \Phi_0)^3] \ . } The curvature of the potential at the
maximum exactly agrees with the open string tachyon mass-squared
$m_T^2 =-1/\alpha'$, obtained as an on-shell condition on the
vertex operator $e^{i Et}$. This serves as a consistency check on
the identification of $\Phi$ with the tachyon field localized on
the D0-branes of \zz.

The open string spectrum also includes a non-dynamical gauge field $A$,
corresponding to the vertex operator $\dot t$.
It enters the covariant derivative in \EQM:
\eqn\cov{
D_x \Phi= \partial_x \Phi - [A,\Phi]
\ .
}
$A$ acts as a lagrange multiplier that projects onto $SU(N)$ singlet
wave functions.

The exact solvability of the model \EQM\ in the singlet
sector relies on the classic fact \BPIZ\
that the $N$ eigenvalues of the matrix $\Phi$ act as free fermions and
can hence be described by Slater determinant wave functions of $N$
variables. The ground state is obtained by filling the first $N$
levels to a Fermi level $-\mu_F$ (as measured from the local maximum of
the potential). To take the double-scaling limit, one sends
$\mu_F \rightarrow 0$, $\beta\rightarrow \infty$, keeping
$\mu = \beta \mu_F$ fixed. The parameter $\mu$ is proportional to
$1/g_s$ and therefore has to be kept large in perturbation theory.
The mass of the unstable D0-brane is the energy
$\mu$ required to move an eigenvalue
from the Fermi level to the top of the potential \mgv, and it scales
correctly as $1/g_s$.

The double-scaling limit zooms in on the local quadratic maximum
of the potential. In this sense,
the $c=1$ matrix model is equivalent to free fermions moving in an
upside down quadratic potential.\foot{
It is sometimes useful to keep in mind that the original potential
includes a
cutoff far from the maximum so that we have a finite number of levels.}
Thus, we have a new example of exact large $N$ duality \mgv.
The $SU(N)$ symmetric
matrix quantum mechanics \EQM\ in an upside down harmonic oscillator
potential, which exactly describes open strings on $N$ D0-branes of
\zz, is dual to Liouville theory coupled to $c=1$ matter which describes
two-dimensional closed string theory together with its D0-branes.

We will be interested in a method for bosonizing the
non-relativistic fermions. Several
closely related ways of doing this are available in the
literature \refs{\dj,\jp}. We will find it convenient to follow the formalism
developed in \refs{\sengupta,\gk} and reviewed in
\igor.
We briefly summarize this method.

The second quantized hamiltonian for a system of free fermions is
\eqn\sec{ {\hat H} \sim \int dy \left\{ {1 \over 2 } {\partial
\Psi^{\dagger}  \over
\partial y }
{\partial \Psi \over \partial y} -{y^2\over 2\alpha'}
\Psi^\dagger \Psi + \mu (\Psi^\dagger \Psi -N)\right\}
\,\,,
}
where $\mu$ is the Lagrange multiplier necessary to fix the
total number of fermions to
equal $N$. As usual, it will
be adjusted so as to equal the Fermi level of the $N$ fermion system.
We may introduce new chiral fermionic variables
$\Psi_L$ and $\Psi_R$ ($t=ix$ is now the Lorentzian time),
\eqn\chiral{
\Psi(y,t) ={ e^{i\mu t}\over { \sqrt{2 v(y)}}} \big[
e^{ -i \int^y  d y' v(y')  + i\pi /4   }\Psi_L(y, t)
+e^{ i \int^y d y' v(y')  - i\pi /4  }\Psi_R(y, t) \big]
\,\,\,,
}
where
\eqn\velocity{
v(y)= {dy\over d\tau}= \sqrt{ {y^2\over \ap} - 2\mu}
}
is the velocity of the classical trajectory
of a particle at the Fermi level.
In terms of the new variables, and using $\tau$
as the spatial coordinate, the Hamiltonian becomes
\eqn\hamil{
\eqalign{ &
\int  d \tau  \biggl[ i \Psi_R^{\dagger}
\partial_{\tau}\Psi_R
-i\Psi_L^{\dagger} \partial_{\tau}\Psi_L + {1 \over 2 v^2}
\big(\partial_{\tau} \Psi_L^{\dagger} \partial_{\tau}\Psi_L +
\partial_{\tau} \Psi_R^{\dagger} \partial_{\tau}\Psi_R \big) \cr
&+ {1 \over 4 } \big( \Psi_L^{\dagger} \Psi_L +\Psi_R^{\dagger}
\Psi_R \big)\left({ v'' \over v^3} - {5 ( v' )^2 \over 2v^4}
\right) \biggr]\ , \cr} }
 where $v' \equiv dv/d\tau$. We notice
that for large $\tau$ (and hence large $v$) the hamiltonian is
approximately relativistic. This large $\tau$ region will be our
primary concern since we are interested in translating to bosonic
variables in the asymptotic future and past.\foot{ One may be
concerned about the region near the turning point of the classical
trajectory where there are singularities in the corrections to
non-relativistic terms. A way to deal with this was suggested in
\refs{\igor,\joe}: instead of wave functions in position space, we
could consider them in momentum space. Due to the special nature
of the upside down harmonic oscillator potential, the hamiltonian
preserves its form, except now the Fermi level is {\it above} the
barrier. After this transformation, the velocity $ v(\tau) = 
\sqrt{2 \mu} \cosh (\tau/\sqrt{\ap}) $ has no zeroes and the range
of $\tau$ may be taken from $-\infty$ to $\infty$. In this
momentum space picture a single branch of Fermi surface has
signals of only one chirality and there are no end-point
singularities. Another way to think about this picture is to
formally send $\mu\rightarrow -\mu$ in the original problem so
that the Fermi level is above the top of the potential \igor.
 Then the D-brane is a hole in the Fermi sea
at precisely the zero energy. Our primary concern will be the
region far from the singularity where either the momentum space or
position space formulation is non-singular. } If we introduce a
cutoff, then the energy levels are approximately given by
\eqn\enerapp{ \epsilon_n  \sim -\mu + n/T + {\delta'(n/T, \mu)
\over T^2} + \cdots }
 where $ 2 \pi T = \oint {d y\over v(y)} =
\oint d\tau $, and $\delta' $ is related to the derivative of the
reflection amplitude from the inverted harmonic oscillator barrier
and from the cutoff barrier. We will be interested in the limit of
large $T$ so the the last term in \enerapp\ does not contribute.
It is important to notice that \enerapp\ is valid even if
$\epsilon_n \sim  \mu $. So we dropped the second term in
\enerapp\ in the limit $T\to \infty$, even though it diverges when
$\epsilon_n \to 0 $ (top of the potential). So, then we can write
the left-moving fermion as in \refs{\sengupta,\gk,\igor}:
 \eqn\psif{
\Psi_L(\tau , t) \sim \int_0^\infty dp e^{ip(\tau+t)} b_p^\dagger
+ \int_{-\infty}^0 dp e^{ip(\tau + t) } d_p \ , }
 in the region
where $y \gg 1$, $\tau\gg 1$. Since here $v\gg 1$, the fermions
are approximately relativistic. We concentrate on the left moving
fermion. There is a similar expression for the right moving
fermions.

The relativistic fermion $\Psi_L$ is bosonized as
 \eqn\boson{
\Psi_L(\tau + t) |0\rangle = e^{ i2\sqrt\pi  \phi_L(\tau + t) }|0>
 }
where $\phi$ is a canonically normalized massless scalar field.
$\phi$ is linearly related to the bulk tachyon field $V$. The
relation is non-local. In Fourier space the relation is just a
momentum dependent phase,
 \eqn\phaserel{ \phi_p = e^{i \delta(p)}V_p . }
This phase is  discussed in more  detail in Appendix A. It has not
been fully derived, as far as we know, but its presence can be
indirectly seen by looking at the Euclidean theory. It is
necessary to include it when we compare the free fermion
amplitudes to the string theory amplitudes.

\subsec{D-brane decay in the matrix model}

If we start with a single fermion near the top of the potential
this fermion will roll down and go to infinity. When it is in the
large $\tau$ region it becomes relativistic and we can bosonize it
using the formulae in \boson . Note then that the state \boson\
agrees precisely with the form of the coherent state that we get
at tree level in string theory \state . However this is not the
precise description. In order to obtain a precise description we
should notice that the free fermion will have some particular
wavefunction $\psi(\tau + t)$. This wavefunction is the one that
results from localizing the eigenvalue on top of the potential. So
a more precise description of the final state would be to write it
as \eqn\wavefn{ \int d\tau \psi(\tau)  \Psi_L (\tau) |0\rangle  =
 \int d\tau \psi (\tau)  e^{i 2\sqrt\pi
\int { dp \over 2\pi\sqrt{ 2  E}} e^{-i p \tau} a^\dagger_p }
|0 \rangle
}
Here $\psi$ is the wavefunction that results from taking a wavefunction
that is initially localized near the top and evolving it to the
asymptotic region and rexpressing it in terms of the relativistic
fermions.

In particular, a single fermion of energy $E$ is
equivalent to
\eqn\energ{
\int d\tau  e^{i E \tau} \Psi_L(\tau) |0\rangle =
 \int d\tau e^{ i E \tau }
e^{i2\sqrt\pi \int { dp \over 2\pi \sqrt{2  E} }
e^{-i p \tau} a^\dagger_p }
|0 \rangle
}
where $a_p$ is the massless boson annihilation operator.
Note that this state precisely agrees with the state that we found
above \state , except for the projection onto a definite energy state.

A more precise discussion has to take into account the fact that
this fermion has to be anti-symmetrized with the fermions forming
the fermi sea. For large $\mu$ these effects are very small.
The full specification of the initial state involves also
saying what the fermions forming the fermi sea are doing. If we
leave them untouched when we add the extra eigenvalue we note that
the matrix model expressions for the closed string emission
agree with the expressions obtained in \llm\ with the so called
``Hartle-Hawking'' contour, which was the one naturally related
to the Euclidean computation.
One surprising consequence of this contour choice is that at tree
level the closed string state seemed to be independent (up to
an overall time delay) of $\tilde \lambda $. We see that this dependence
will come in when we include the wavefunction of the initial
state, as in \wavefn .

Sen's boundary state with parameter $\tilde \lambda$  \SenNU\
corresponds,
 in the matrix model, to an eigenvalue that starts at  the position
\eqn\position{ y = -\sqrt{2\alpha'\mu} \sin \pi \tilde \lambda }
 Then the
energy of the state is given by $E = \mu \cos^2 (\pi \tilde \lambda)$,
as in \SenNU .
 The time delay in the
classical evolution of this trajectory relative to the classical
trajectory at the Fermi level is \eqn\timedelay{ \Delta t =
\log(\sin \pi \tilde \lambda) } It is easy to check that this is
precisely the time delay that appears as a phase in \timepart .

\subsec{D-branes in the Euclidean 2d String}

Let us also try to give matrix model interpretation to D-branes of
the Euclidean 2-d string theory localized at large $\phi$. Putting
the Neumann boundary condition on the field $X$ once again
corresponds to placing an eigenvalue at the top of the upside down
potential. It is well-known \refs{\Callan,\larus} that one can
turn on an exactly marginal boundary operator \eqn\exactmar{
\lambda \oint d\sigma \cos (X/\sqrt{\ap}) \ ,} which interpolates
between the Neumann and the Dirichlet boundary conditions on $X$.
We believe that turning on this potential corresponds to an
eigenvalue in the matrix model executing classical Euclidean
motion in the forbidden region \eqn\exactmot{ y(x) = \sqrt{2
\alpha'\mu} \sin \pi \tilde \lambda
 \cos (x/\sqrt{\ap}) \ .
}
The value $\tilde \lambda=1/2$, corresponding to the Dirichlet
boundary condition on $X$, is dual to the eigenvalue trajectory at
the Fermi level. Thus, the D-instanton is indeed related to the
eigenvalue tunnelling at the Fermi level, as speculated in \Polch.

There is an interesting subtlety, however. In the matrix model the
energy of an eigenvalue at the maximum is $\mu$, while the action
of the tunnelling trajectory is $  \pi \mu \sqrt{\alpha'} $. The
tunnelling trajectory has to be interpreted as `half' of a
D-instanton.\foot{For a recent discussion of half-instantons and
merons, see \DrukkerWX.}
 The D-instanton corresponds to tunnelling from
the filled region to the unfilled, and then tunnelling back (i.e.
the bounce). Then the D-instanton action $2 \pi \mu \sqrt{\alpha'}
$ is related to the tension of the D0 brane by \eqn\actins{ {
T_{-1} \over T_0} =  2 \pi \sqrt{\alpha'} \ .} This agrees with
the descent relation following from the boundary state formalism.
This relation is completely determined by the Neumann and
Dirichlet boundary states of the field $X$ and does not involve
the knowledge of the Liouville boundary state.

\newsec{Discussion}

Notice that the divergences in the energy and norm of the state
that we had in the tree level result \abs\ are absent from the all
orders result \wavefn . We also clearly see that the total energy
emitted is obviously equal to the energy of the initial state
characterized by $\psi$. In this example the divergence of the
tree level answer is regularized by considering the quantum
mechanics of the tachyon field. So it looks like a one loop effect
in the open string channel. In this case the fermions are
completely free, so it is hard to extract a general lesson for the
26 dimensional bosonic string. One lesson is that taking into
account the quantum mechanics of the open strings is crucial for
finding a regulated expression. Another general lesson is that the
wavefunction of the open string modes gets imprinted in the final
wave function for the closed strings. In this completely solvable
example we find that the ``tachyon matter'' state introduced in
\SenNU\ is just the state of  closed strings that the brane decays
into.

Another interesting issue is the relation between the tachyon potentials
discussed in the context of string field theory
\refs{\SenNX,\GerasimovZP,\gregk}
 to
the simple inverted quadratic potential we find here.
We think that the relation is the following. The
string field theory potential is only part of the parabola.
It is  the region above the fermi sea on the left
side and the whole parabola on the right side (the unfilled side).
This agrees with two
qualitative features. One is that at a finite distance there is
a minimum corresponding to the closed string vacuum, this
corresponds to placing an eigenvalue at the fermi surface. On the
``wrong'' side the potential is unbounded below.

Let us also comment on the D1-branes obtained from the extended
Liouville boundary state of \refs{\fzz,\teschner} multiplied by
the Neumann boundary state for the (Euclidean) time. We believe
that a dual large $N$ description of the 2-d string theory in
presence of such a space-time filling brane is given by \eqn\VEQM{
Z= \int D^{N^2}\Phi(x) D^{2N} V(x) e^{-\beta S} \ ,} \eqn\VEQMS{
S= \int_{-\infty }^{\infty }dx~\biggl [\Tr \left (\half (D_x \Phi
)^2+ U(\Phi)\right )+ \half (D_x V)^\dagger D_x V + \half m^2
V^\dagger V + g V^\dagger \Phi V\biggr ]\ , } where $V$ is a
complex vector. Integrating over $V$ inserts into random surfaces
dynamical boundaries that can wander in the time direction (models
of this type were considered in \refs{\yang,\min,\kk}). By tuning
$m^2$ and $g$ together with $\beta$, we expect to find a scaling
model with independent parameters $\mu_b$ and $\mu$ as in \bound.
The massless open string `tachyon' is described by operators $\int
dx e^{i q x} V^\dagger V(x)$.

If we consider a model without the kinetic term for $V$, then
integration over $V$ generates boundaries localized in the time
direction, with subsequent averaging over possible locations of
the boundary. Such boundaries with Dirichlet boundary conditions
on the time coordinate are the macroscopic loops which were
studied in \refs{\MooreIR,\nati}.

We should also emphasize another important distinction between the
D0-brane and the macroscopic loop.   The former is localized in
$\phi$, while the latter extends all the way to the spatial
boundary at $\phi \to -\infty$. Therefore, the macroscopic loop is
an observable in the theory which is associated with a change in
the boundary conditions. It differs from the localized D0-brane
which corresponds to a dynamical excitation of the system.

It is important to distinguish the eigenvalue $y$ from the
Liouville coordinate $\phi$.  The spaces parameterized by them are
related to each other through a nonlocal transform
\refs{\nati,\PolchinskiJP}.  The time-dependence of an eigenvalue
(the open string tachyon) does not describe motion of a D0-brane
in $\phi$; rather, it describes the decay of a D0-brane localized
at large $\phi$ \SenNU. The time dependence in the $\phi$
description arises only from the boundary interaction \inter.

\bigskip

{\bf Acknowledgments}

We would like to thank N. Itzhaki, J. McGreevy, A.M. Polyakov and
H. Verlinde for useful discussions. The research of JM and NS is
supported in part by DOE grant DE-FG02-90ER40542. The research of
IRK was supported in part by the National Science Foundation
Grants No. PHY-9802484 and PHY-0243680. Any opinions, findings,
and conclusions or recommendations expressed in this material are
those of the authors and do not necessarily reflect the views of
the National Science Foundation.

\appendix{A}{Leg Factors}

In this appendix we discuss a bit more the phase \phaserel\
 present in the
relation between the string theory computations and the matrix
model computations. These leg factors are typical in relations
between closed string amplitudes and their origin was discussed,
for example, in \refs{\nati,\igor}. Let us first discuss the
Euclidean matrix model. It has finite boundary operator (whose
$l\rightarrow 0$ limit is the puncture operator)
 \eqn\fbound{ O(l,
q)=\int dx e^{iqx} \Tr e^{-l\Phi(x)} }
 which is translated into
 \eqn\transl{ {1\over 2} \int dx e^{iqx} \int_0^\infty d \tau e^{-l
y (\tau)} : \Psi^{\dagger}_L \Psi_L +\Psi^{\dagger}_R \Psi_R(\tau,
x): \ , } where $y (\tau)$ is the classical trajectory at the
Fermi level. Upon bosonization, we find \eqn\bosonp{ O(l, q)\sim
\int dx e^{iqx} \int d \tau e^{-l y(\tau)}\partial_\tau X\sim
i\int_{-\infty}^\infty dk F(k, l) k \tilde X(q, k) } where
 \eqn\transform{ F(k, l)=\int_0^\infty d\tau
e^{-l\sqrt{2\ap\mu}\cosh (\ap\tau)}\cos(k\tau) \ . }
In working
with the Euclidean matrix model, there is freedom in multiplying
puncture operator $O(l\rightarrow 0, q)$ corresponding to tachyon
of momentum $q$ by a smooth function of $q$ \igor. The
normalization may be chosen in such a way that, in the correlation
function, each operator is accompanied by a leg factor containing
all the poles due to the discrete states:
 \eqn\leg{ \mu^{\sqrt
{\ap} |q|/2} {\Gamma (-\sqrt{\ap} |q|)\over \Gamma (\sqrt{\ap}
|q|)} \ . }
 We have picked the numerator so that it reproduces the
poles present in the relation. The special property of this choice
of normalization is that after continuation to Lorentzian
signature, $X\rightarrow it$, the leg factor corresponding to
operator $ e^{\pm i Pt} e^{(2 \ap^{-1/2}+ iP)\phi}$ becomes a pure
phase
 \eqn\phase{ \mu^{ - {i\over 2}\sqrt{\ap} P} { \Gamma( i
\sqrt{\ap} P) \over \Gamma(- i \sqrt{\ap} P )} \ . }
 These phases
are indeed present in all Liouville theory calculations and, in
particular, in the disk 1-point function \abs. However, these
phases do not appear in the S-matrix obtained using Lorentzian
semiclassical free fermions with Polchinski's methods \joe.
Therefore, to compare free fermion calculations with Liouville
theory, the phases have to be added ``by hand.'' As we explained
above, one justification for these phases is to start in the
Euclidean signature, where the leg factors are certain observable
(they contain poles due to the discrete states), and later
continue to Lorentzian signature.

\appendix{B}{ Matching of coefficients}

In this Appendix we match the coefficients between the
free fermion computation and the Liouville computation.

Let us start with the matrix model answer
\eqn\asmm{
\Psi_L|0\rangle  = e^{ i 2\sqrt{\pi} \phi_L}|0\rangle
=
 e^{ i 2\sqrt{\pi} \int_0^\infty
 { dp \over 2 \pi } { 1 \over \sqrt{2 E} } a^\dagger_p }|0\rangle
}
 where $\phi$ is a canonically normalized scalar field, and
similarly the $a_p^\dagger$ are the creation operators canonically
normalized, with the usual continuum normalization ( $[a_p,
a_{p'}^\dagger] = 2 \pi \delta(p-p') $). In the string theory we
find that the amplitude for emitting on shell strings is given by
\eqn\emmampl{ {\cal A} =  {\cal N} e^{i \delta} } where $e^{i
\delta}$ is an unimportant phase. Our goal is to show that the
normalization factor is ${\cal N}= 2\sqrt\pi$ in agreement with
the matrix model.

For this purpose we note that formally the emission amplitude
is related to the imaginary part of the one loop Feynman amplitude
through
\eqn\relation{
 2 Im(Z_1) = \int_0^\infty { dp \over 2 \pi } { |{\cal A}|^2 \over 2 E(p) }
} We start with the one loop amplitude for open strings for the
brane with Neuman boundary conditions in the time direction. Using
the character for the $\chi_{(1,1)}$ representation of \zz, this
is given by \eqn\parfn{ Z_1 = \int_0^\infty { dt \over 2 t }
Tr_{open} e^{ - 2 \pi t L_0} = V_t \int_0^\infty  {dt \over 2 t}
\int_{-\infty}^\infty {d k_0 \over 2 \pi } e^{ -2 \pi t \alpha'
k_0^2 }  ( e^{ + 2 \pi t } -1) } where $V_t$ is the length of the
time direction. This partition function shows only two open string
states, the tachyon and the $A_0$ gauge constraint (which gives
the $-1$). We rewrite it in the closed string channel as ($\alpha'
=1$) \eqn\closed{ Z_1 = V_t \int_0^\infty ds \int_0^\infty { dp
\over 2 \pi } (\sinh \pi p)^2 e^{ - 2 \pi s {p^2 \over 4} } } The
momentum integral runs over half the real axis because the closed
string states labeled by $p$ and $-p$ are the same (up to a phase
which is the reflection amplitude). If we consider instead the
decaying brane we should substitute \llm\ \eqn\subst{ V_t = \int {
dk_0 \over 2 \pi}  ( 2 \pi \delta(k_0) )^2 \to \int { d k_0 \over
2 \pi } \left( {\pi \over \sinh \pi k_0 } \right)^2 e^{ -2 \pi s
\alpha' { -k_0^2 \over 4} } } Inserting this in \closed\ and
(formally) integrating over $s$
 find that we get the expression
\eqn\onelop{
Z_1 = { 2 \over \pi } \int_0^\infty { dp \over 2 \pi }
\int_{-\infty}^\infty { d k_0 \over
2 \pi } \left( { \pi \sinh \pi p \over \sinh \pi k_0 } \right)^2
{ 1 \over - k_0^2 + p^2  - i \epsilon }
}
The imaginary part comes from the two poles at $k_0 = \pm p$ and
gives
\eqn\gives{
Im\left[ \int { d k_0  \over 2 \pi }
{  1 \over - k_0^2 + p^2  - i \epsilon } f(k_0)\right] =  { f(E(p)) \over
2 E(p) } ~;~~~~~~~~~E(p) = p
}
So we find
\eqn\fingin{
2 Im(Z_1) = { 4  \pi } \int_0^\infty
{d p \over 2 \pi } { 1 \over 2 E(p)}
}
Comparing this with \relation\ we find
that
\eqn\finalform{
{\cal A} =  2 \sqrt{ \pi} e^{ i \delta}
}

\listrefs

\bye